\documentclass[RNAAS]{aastex62}

\graphicspath{{./}{figures/}}
\usepackage{graphicx}
\usepackage{todonotes}
\begin{document}
\title{The Diffuse Supernova Neutrino Background }

\correspondingauthor{Samalka Anandagoda}
\email{iananda@g.clemson.edu}

\author[0000-0002-5490-2689]{Samalka Anandagoda}
\affiliation{Clemson University, Department of Physics $\&$ Astronomy, Clemson, SC 29634-0978}

\author{Dieter H. Hartmann}
\affiliation{Clemson University, Department of Physics $\&$ Astronomy, Clemson, SC 29634-0978}
\author{Marco Ajello}
\affiliation{Clemson University, Department of Physics $\&$ Astronomy, Clemson, SC 29634-0978}

\author{Abhishek Desai}
\affiliation{Clemson University, Department of Physics $\&$ Astronomy, Clemson, SC 29634-0978}

%% Note that RNAAS manuscripts DO NOT have abstracts.
%\keywords{diffuse radiation --- neutrinos ---  supernovae: general}

%% Start the main body of the article. If no sections in the 
%% research note leave the \section call blank to make the title.
\section{}

\vspace{-0.3cm}
Multi-Messenger Astrophysics began with the detection of 25 neutrinos from SN1987A \citep{hirata87,bionta87,alex87}. Their distribution in time and energy confirmed the paradigm of the formation of a hot proto-neutron star (PNS) in a core collapse Supernova (ccSN) \cite{JANKA2007}.
To reveal the dynamics of these environments, time-resolved spectroscopy with high count rates is required, which is possible when a ccSN occurs in the Milky Way (MW), at a low rate of $\leq$3 per century \citep{MWccsn}. However, advances in technology point to ccSN $\nu$-detections in the local volume (D$_{\rm{LV}}=10\rm{Mpc}$), at a rate of $\sim$ 1 yr$^{-1}$ \citep{Anandagoda2019,Ando2005}, and existing detectors offer an independent route to probe ccSNe through the Diffuse Supernova Neutrino Background (DSNB; e.g., \citeauthor{beacom10}, \citeyear{beacom10}). The DSNB in the MeV regime represents the cumulative cosmic neutrino emission, predominantly due to ccSNe.
Estimates of the DSNB flux found values of 0.2 $\pm$ 0.1\,cm$^{-2}$ s$^{-1}$ \citep{Hartmann-woosley97},
0.14-0.46\,cm$^{-2}$ s$^{-1}$ (E$_\nu$ $>$19.3 MeV) \citep{Ando_2004} and $\sim$ 0.24\,cm$^{-2}$ s$^{-1}$ in the energy range 19.3-27.3 MeV \citep{Horiuchi09}. Ongoing observations by Super-Kamiokande report an upper limit of 2.8-3.1 $\overline{\nu}_e$\,cm$^{-2}$s$^{-1}$ (90$\%$ C.L) for E$_\nu$ $>$ 17.3\,MeV \citep{SK-experiment}. \\

We estimate the DSNB flux for different Star Formation Rate Density (SFRD) models and find that the DSNB 
can be used to estimate the SFRD using a method similar to the one employed for the Extragalactic Background Light (EBL) by the \cite{2018Science}.\\

The DSNB (in the energy-window (E$_0$,E$_1$) \footnote{Boundaries set by competing solar and atmospheric backgrounds are chosen to be $\rm{E_0}=19.3$\,MeV and $\rm{E_1}=40$\,MeV.}), for a given SFRD, $\psi(z)$ in units of $\rm{s^{-1}\,cm^{-2}\,MeV^{-1}}$, is given by

\begin{equation}
\rm{F_\nu} =c\,\rm{\int_{E_0}^{E_1} \int_0^\infty dE_\nu\,dz \ C(z)\times R_{SN}(z) \times  \frac{dN_\nu}{dE_{\nu}}(E_{\nu}') } 
\label{eq:dsnb}
\end{equation}

where the ccSN rate is $\rm{R_{SN} = (\psi(z) / <m>) \times f_{SN} \ year^{-1}\, Mpc^{-3}}$. The average star mass, $\rm{<m>}$,
and the fraction of stars that produce a ccSN, f$_{\rm SN}$, depend on the Initial Mass Function (IMF). We adopt a flat $\Lambda$CDM cosmology where the cosmological factor in Eq.(\ref{eq:dsnb}) is  $\rm{C(z})$ = 1/(H$_0$E(z)) where ${\rm E(z) = (\Omega_m (1+z)^3 + \Omega_\Lambda})^{1/2}$. See \cite{Ando_2004}.\\

To model the emergent neutrino spectrum of a single ccSN, given by $\rm{\frac{dN_\nu}{dE_{\nu}}(E_{\nu}')}$, in Eq.(\ref{eq:dsnb}), we follow \cite{Hartmann-woosley97,beacom10}.
Assuming a Fermi Dirac (FD) distribution of temperature T, zero chemical potential with an average neutrinosphere radius, the $\overline{\nu}_e$ spectrum is, 

\begin{equation}
\rm{\frac{dN_\nu}{dE_{\nu}}(E_{\nu}')} = E_{\nu,tot} \ \frac{120}{7 \pi^4}\ E{'}_\nu^2\ [T^4(e^{{E'_\nu}/{T}}+1)]^{-1} \qquad  (\nu\,MeV^{-1})
\label{eq:dnde}
\end{equation}

where, E$_\nu'$=(1+z)E$_\nu$ and the total energy released per flavor ($\rm{E_{\nu,tot}}$) is assumed to be 1/6 of the total Binding energy (BE) 
\citep{Neutrino-flavor}. BE is from equation 36 of \cite{Lattimer} assuming a neutron star mass and radius of $\rm M_{NS}={1.4\ M_\odot}$ and  $\rm{R_{NS}=11\ km}$. \\

The SFRD(z) is determined through various tracers (UV,IR continuum, H$\alpha$ line) and source counts as a function of redshift eg: \citep{madau14}. Therefore, the SFRD(z) depends on cosmology and the IMF. Modifying the SFRD for different cosmological parameters introduces a factor, H$_0$ E(z) \citep{Porciani_2001} that cancels C(z) appearing in Eq.(\ref{eq:dsnb}), rendering it independent of cosmological changes in SFRD models  \citep[see][]{Ando_2004}.\\

\begin{figure}[htb]
\vspace{-0.34cm}
\centering
\includegraphics[width=1.0\textwidth]{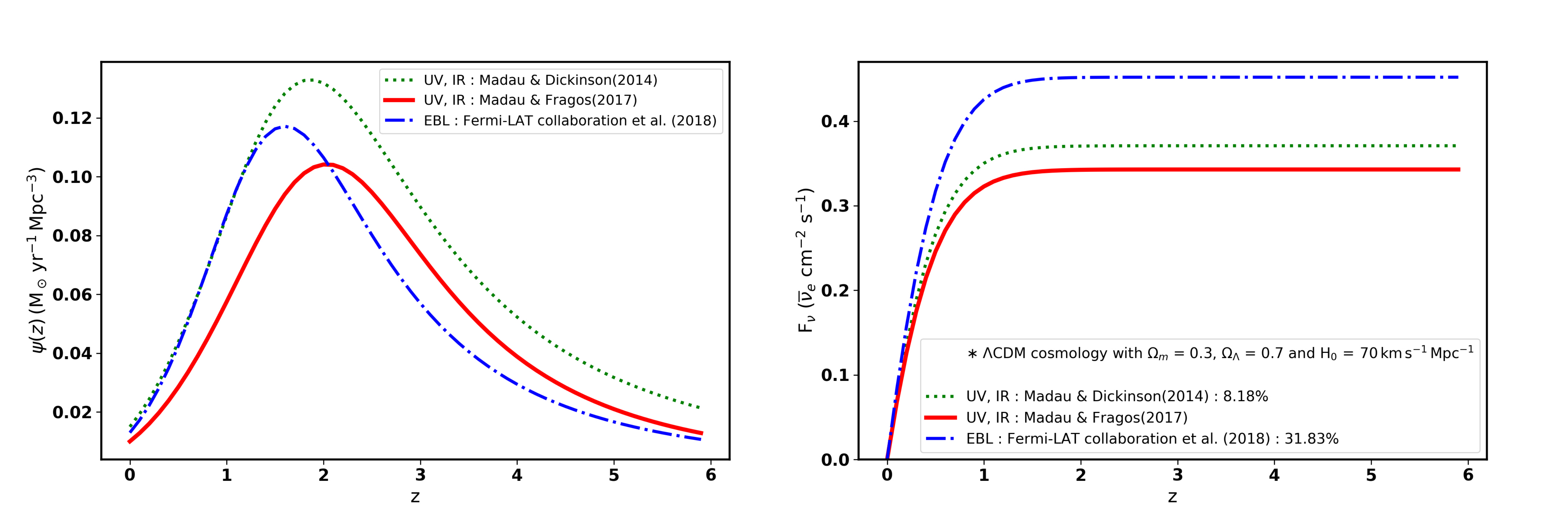}
\caption{Variation of SFRD models with redshift (left) and SFRD model dependence of the DSNB flux (right), keeping all other parameters at their standard values. %$\rm{h_x}$ is defined as h$_{\rm{x}}$=H$_0$/x km s$^{-1}$ Mpc$^{-1}$=1.
The Percentage increase of the DSNB flux for different SFRD models relative to the benchmark (red) is indicated.}
\label{fig}
\vspace{-0.25cm}
\end{figure}

We use three SFRD models: 1.\cite{madau14}, 2. \cite{Madau_2017} and 3. \cite{2018Science}\footnote{Derived using EBL intensity measurements}. To compute the DSNB flux (see Figure~\ref{fig}:right), C(z) is established via H$_0 =70$ $\rm{km\,s^{-1}\,Mpc^{-1}}$ and $\Omega_{\rm{m}} = 0.3$ and the antielectron neutrino spectrum is given by Eq.(\ref{eq:dnde}) with T$ =4.76$\,MeV
\citep{beacom10}. The IMFs associated with these models are \cite{Salpeter_1955}, \cite{Kroupa_2001} and \cite{Chabrier_2003}, respectively. As a benchmark, we select the SFRD(z) of \cite{Madau_2017} given by 

\begin{equation}
\psi(z)= \frac{(1+z)^{\rm{a}}}{1+(\frac{1+z}{\rm{b}})^{\rm{c}}}\times \rm{d} \qquad \qquad (\rm{M_\odot \,yr^{-1}\,Mpc^{-3}})
\label{eq:sfrd}
\end{equation}

with parameters: a $=2.6$, b $=3.2$, c $=6.2$ and d $=0.01$. The SFRD(z) model estimated by \cite{2018Science} has parameters: a $=2.99$, b $=2.63$, c $=6.19$, d $=0.013$ (Figure~\ref{fig}:left:blue). \\

Less than 1$\%$ of the DSNB is due to star formation above z $\sim 2$ since star formation decreases for higher redshifts. Thus uncertainties in the high-z SFRD are not crucial (see Figure~\ref{fig}:right). 

We assume that the mean ccSN is represented by SN1987A (in regards to its neutrino signature), but variations occur due to changing core collapse outcomes and deviations from a FD spectrum with progenitor mass.
The DSNB flux increases significantly $\approx$ 32$\%$ for the SFRD from \cite{2018Science} relative to the benchmark, F$_0$ of 0.34 cm$^{-2}$ s$^{-1}$ in the energy window(19.3-40 MeV). Thus detection of the DSNB with advanced detectors like gadolinium enhanced Super-Kamiokande \citep{Beacom_Vagins} can be utilized to not only probe ccSNe physics \citep{Mathews_2019} but also as a tool to estimate the SFRD.
\acknowledgments

We thank Prof. B. S. Meyer for stimulating discussions.

\end{document}